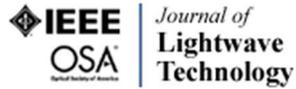

# Guided Plasmon Modes of Elliptical Cross Section Silver Nanoridges

# Guided Plasmon Modes of Elliptical Cross Section Silver Nanoridges

Zeyu Pan, Junpeng Guo, Richard Soref, and Walter Buchwald

*Abstract*—Propagating two-dimensional plasmon modes guided by elliptical cross section silver nanoridges are investigated in this paper. Mode field profiles, dispersion curves, propagation distances, and figure-of-merits of the plasmon ridge modes are calculated for various elliptical cross sections. It is found that an elliptical cross section metal nanoridge, if designed properly, can support a well-confined plasmon mode with a longer propagation distance and a higher figure-of-merit than the flat-top nanoridge plasmon waveguide of the same width. The optimal ridge waveguide cross section is obtained when the elliptical ridge top has a semicircular cross section. When the curvature of the elliptical nanoridge is large, the mode approaches the tightly confined wedge plasmon mode.

## I. INTRODUCTION

Surface plasmons are free electron density oscillations on the metal-dielectric interfaces, and can propagate along the metal-dielectric boundaries in the form of surface plasmon waves with tightly confined sub-wavelength modes [1-3]. Surface plasmon modes in various waveguide structures, such as thin metal films [4-8], finite width thin film metal stripes and metal wires [9-17], trenches in the metal surface [18-29], metal dielectric layer structures [30-39], dielectric-loaded metal films [40-46] and metal wedges [25, 26, 47-52], have been extensively investigated. However, although these waveguides support tightly confined plasmon modes, they also suffer from high attenuation, which severely limits their applications. The quest for strong mode confinement and low loss waveguides has inspired the research efforts on various waveguide structures.

Manuscript received February 09, 2012. This work was supported in part by the National Science Foundation (NSF) through the award NSF-0814103 and the National Aeronautics and Space Administration (NASA) through the grant NNX07 AL52A. J. Guo acknowledges the support from the ASEE-Air Force Office of Scientific Research (AFOSR) Summer Faculty Fellowship Program.

Z. Pan is with the Department of Electrical and Computer Engineering, University of Alabama in Huntsville, Huntsville, Alabama 35899, USA (e-mail: zp0002@uah.edu).

J. Guo is with the Department of Electrical and Computer Engineering, University of Alabama in Huntsville, Huntsville, Alabama 35899, USA (e-mail: guoj@uah.edu).

R. Soref is with the Department of Physics, University of Massachusetts at Boston, Boston, Massachusetts 02125, USA (e-mail: soref@rcn.com).

W. Buchwald is with the Solid State Scientific Corporation, Hollis, New Hampshire 03049, USA (e-mail: walter.buchwald@solidstatescientific.com).

Color versions of one or more of the figures in this paper are available online at http://ieeexplore.ieee.org.

Recently, experimental investigation concerning a gold nanoridge, fabricated with the focused ion beam milling technique [52], confirms that metal nanoridges can also support propagating surface plasmon modes. The 2D confined plasmon modes of flat-top silver nanoridges have been investigated recently [53].

This paper presents comprehensive numerical investigations of the plasmon modes supported by elliptical cross section silver nanoridges. We calculate the mode field distributions, mode indices, dispersion curves, propagation distances, mode sizes, and the figure-of-merits of these silver nanoridges with various surface curvatures as well as their dependence on the wavelength and the ridge width. The motivation for investigating such elliptical cross section silver nanoridge waveguides is to find an optimal geometry that provides longer propagation distance and higher figure-of-merit. In addition, as will be shown, elliptical cross section nanoridge waveguides provide more uniform distributions of the surface plasmon mode fields when compared to the flat-top and wedge plasmon waveguides. This feature is important in applications requiring an expanded interaction region such as in chemical and biological sensing.

## II. ELLIPTICAL CROSS SECTION SILVER NANORIDGE PLASMON WAVEGUIDES

The 3D view and cross section of such an elliptical silver nanoridge are shown in Fig. 1(a) and (b), respectively, where the nano-scale metal ridge is extended in the *z*-direction with its width (*w*) in the *x*-direction and the half ellipse height (*h*) in the *y*-direction. The plasmon wave propagates along the ridge top in the *z*-direction. We assume the height of the ridge is sufficiently large so that the substrate does not influence the ridge plasmon mode. Our previous work on the plasmon modes supported by the flat-top silver nanoridges [53] has found that a ridge width of 120 *nm* produces an optimal compromise between propagation distance and mode confinement. In this study, we thus fix the ridge width at 120 *nm*, for all elliptical cross section nanoridge waveguides investigated.

To describe the elliptical cross section, we use the curvature (*κ*), instead of the ellipse height (*h*). The curvature is the reciprocal of the radius of curvature. The radius of curvature (*R*) of the elliptical cross section ridge at the top of the ridge is $R=w^2/4h$ [54]. Thus, the curvature at the top of the ridge is $\kappa=1/R=4h/w^2$.

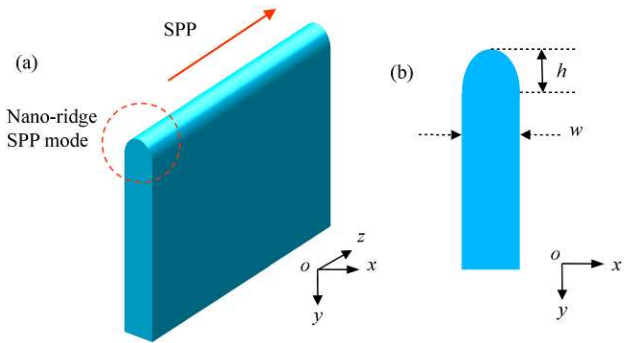

Fig. 1. (a) 3D view of the elliptical nanoridge plasmon waveguide. (b) Cross section of the elliptical nanoridge plasmon waveguide.

We consider the situation where the surrounding medium of the ridge waveguides is air ($\varepsilon_d = 1.0$), and the ridges are made of silver with its electric permittivity $\varepsilon_m = -127.5 - 5.3j$ at the telecommunication wavelength of 1.55 $\mu m$ [55]. For the flat-top ridge with a width of 120 $nm$, the mode effective index and the attenuation coefficient are found to be $n_{eff} = 1.018 - 0.00074j$ and 261.87 $dB/cm$, respectively. While this study investigates silver as the ridge material, the analysis can obviously be extended to other types of high electron density materials, such as heavily doped semiconductors [56-58] albeit at different wavelengths. Fig. 2 (a)-(c) are the mode profiles of the three electric field components ($E_x$, $E_y$, $E_z$) of the 120 $nm$ wide elliptical silver nanoridge plasmon waveguide at the 1.55 $\mu m$ wavelength. Fig. 2 (d)-(f) are the mode profiles of the three magnetic field components ($H_x$, $H_y$, $H_z$) of the elliptical nanoridge waveguide at the same wavelength. The major components of the electric field are the transverse components $E_x$ and $E_y$, as expected. The longitudinal component of the electric field $E_z$ is seen to be roughly three orders of magnitude less than the transverse components. The major magnetic field components are also the transverse components $H_x$ and $H_y$, while the longitudinal component of the magnetic field $H_z$ is also seen to be three orders of magnitude less than the transverse components. Based on these results, the elliptical cross section nanoridge plasmon modes can be considered as a quasi-transverse electromagnetic (TEM) mode. From Fig. 2, we can see that the $E_y$ mode profile and $H_x$ mode profile are symmetrical with respect to the center of the metal ridge (i.e. $x=0$ plane), while the $E_x$ and $H_y$ profiles are anti-symmetrical with respect to the $x=0$ plane as expected, for such a plasmon mode propagating along the top surface of an elliptical metal ridge.

We calculated the electric field intensity distributions of the fundamental plasmon propagating modes supported by elliptical nanoridges of different curvatures at the wavelength of 1.55 $\mu m$ with width of the nanoridges set to 120 $nm$ in all cases. Fig. 3(a) shows that the electric field intensity distribution of the flat-top nanoridge has two hot spots located at each corner of the ridge, which can be considered as a hybrid mode of two 90° wedge plasmon modes first described in Ref. [53]. Fig. 3(b-f) is the electric field intensity distribution of the elliptical nanoridge plasmon mode with the curvature of 15 $\mu m^{-1}$, 30 $\mu m^{-1}$, 70 $\mu m^{-1}$, 110 $\mu m^{-1}$, and 150 $\mu m^{-1}$, respectively. It can be seen that for the elliptical nanoridge with a curvature between 9 and 18 $\mu m^{-1}$, the mode spans the entire ridge. As the curvature increases, the mode progressively converges towards the tip of the ellipse with increasing mode confinement. As the curvature increases to above 18 $\mu m^{-1}$, the mode converges to only one hot spot at the tip, which can be considered as a small angle wedge mode. We also find that the mode energy is distributed over the entire ridge when the curvature is small. In this case, the elliptical nanoridge mode is seen to be insensitive to the change of curvature. However, once the curvature becomes greater than 18 $\mu m^{-1}$, the plasmon mode is seen to be concentrated at the tip of the elliptical ridge, and is seen to rapidly intensify with the increasing curvature.

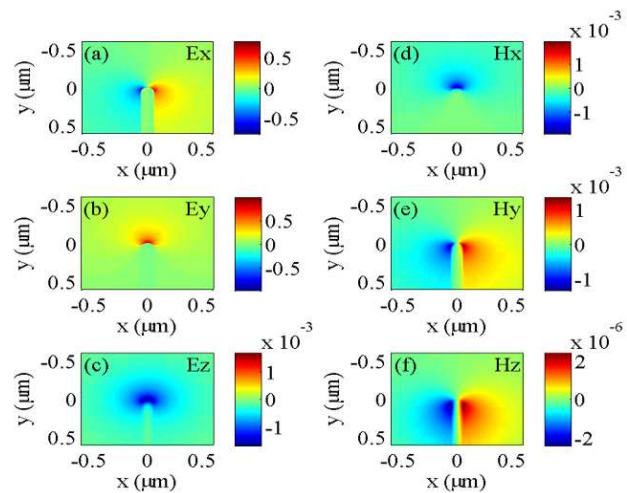

Fig. 2. (a)-(c) are the electric field ($E_x$, $E_y$, $E_z$) mode profiles of an elliptical silver nanoridge with $\kappa = 16.67$ $\mu m^{-1}$ ($h = 60$ $nm$) at 1.55 $\mu m$ wavelength; (d)-(f) are the magnetic field ($H_x$, $H_y$, $H_z$) mode profiles of the elliptical silver nanoridge with $\kappa = 16.67$ $\mu m^{-1}$ ($h = 60$ $nm$) at 1.55 $\mu m$ wavelength.

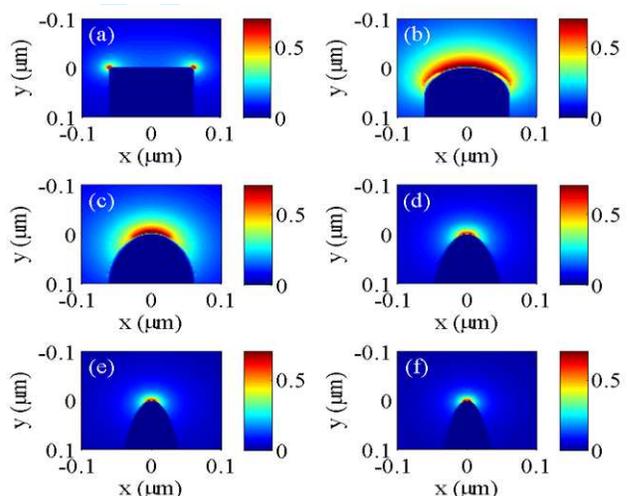

Fig. 3. Electric field intensity distributions of the 120 $nm$ wide elliptical nanoridge plasmon waveguide with curvature equal to (a) 0 $\mu m^{-1}$, (b) 15 $\mu m^{-1}$, (c) 30 $\mu m^{-1}$, (d) 70 $\mu m^{-1}$, (e) 110 $\mu m^{-1}$, and (f) 150 $\mu m^{-1}$ at 1.55 $\mu m$ wavelength.

The mode dispersion curves for the elliptical silver nanoridges of different curvatures are calculated and shown in Fig. 4. The black solid line is the light line in air, and the black dashed line is the plasmon dispersion curve associated with the silver-air flat interface. In this figure, for all the curvatures, as the frequency increases, the plasmon mode dispersion curves drift away from the light line in air, suggesting slower group velocity and tighter mode confinement. This effect is far more pronounced for ridges with larger curvature. This is in contrast to the effects of the curvature on the dispersion curve, which is more easily seen in the inset of Fig. 4. Here, as the curvature increases from zero (i.e. the flat-top nanoridge), the dispersion curve first moves toward the light line (i.e. to the left side of the flat-top nanoridge dispersion curve in the inset), indicating a reduction in mode confinement and propagation attenuation. Once the curvature increases greater than 16.67 $\mu m^{-1}$, the dispersion curve moves away from the light line and eventually falls in the right side of the zero curvature case, indicating the increase of both the mode confinement and propagation attenuation. The red line is the dispersion curve of the flat-top nanoridge waveguide. The blue line is the dispersion curve of the nanoridge waveguide with 15 $\mu m^{-1}$ curvature. The green line is the dispersion curve of the nanoridge waveguide with 30 $\mu m^{-1}$ curvature. The magenta and cyan dashed lines are the dispersion curves of the nanoridge waveguides with the curvature of 90 $\mu m^{-1}$ and 150 $\mu m^{-1}$, respectively. It is also seen that the dispersion curves of ridges with curvatures between 9 and 18 $\mu m^{-1}$ are relatively insensitive to curvature changes. While beyond a curvature of 18 $\mu m^{-1}$, the mode intensity is more concentrated to the tip of the elliptical ridge. The dispersion curve moves rapidly away from the light line when the curvature of the elliptical ridge continues to increase, and the ridge plasmon mode gradually approaches a wedge plasmon mode.

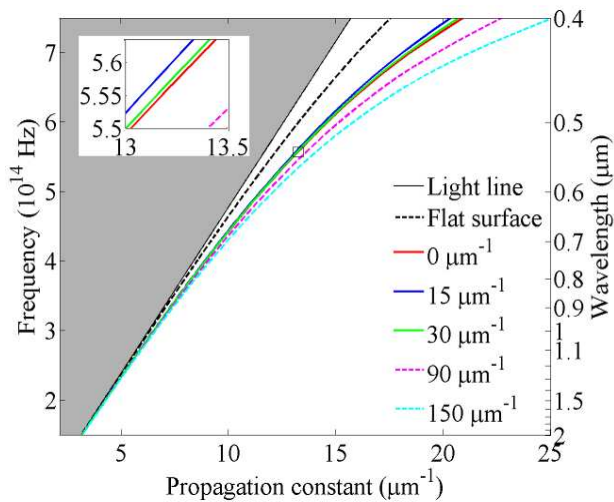

Fig. 4. Dispersion curves of the silver elliptical nanoridge plasmon waveguides of different curvatures and the comparison with the dispersion curve of the silver-air flat plasmon mode.

The real and imaginary parts of the elliptical nanoridge mode index versus the wavelength for several different curvatures ($\kappa$ = 0 $\mu m^{-1}$, 15 $\mu m^{-1}$, 30 $\mu m^{-1}$, 90 $\mu m^{-1}$, and 150 $\mu m^{-1}$) are plotted in Fig. 5, where the black dashed lines are those of the silver-air flat plasmon mode. As the wavelength increases, both the real and imaginary parts of the elliptical nanoridge mode index decrease, indicating a reduction in the mode confinement as well as the propagation loss. We can also see when the curvature increases from the zero, both the real and imaginary part of the elliptical nanoridge mode index first reduce, and further increase of the curvature increases those of the elliptical nanoridge mode index, which is consistent with the results shown in Figs. 3 and 4.

The real and imaginary parts of the mode index versus the curvature at 1.55 $\mu m$ wavelength are shown in Fig. 6. Interestingly, both the real and imaginary parts of the elliptical nanoridge mode index are first seen to decrease with an initial increase in the elliptical curvature, indicating a reduction in propagation attenuation and mode confinement, however, after reaching minimum values at 16.67 $\mu m^{-1}$, rapid increase in both real and imaginary parts of the mode index is observed, which corresponds to a quick rise in both attenuation and confinement. It is interesting to note that considering the 120 $nm$ ridge width here, a semi-circle of 60 $nm$ radius would have a curvature of 16.67 $\mu m^{-1}$. This suggests that a semicircular geometry is the optimal for reduced propagation loss and increased mode confinement.

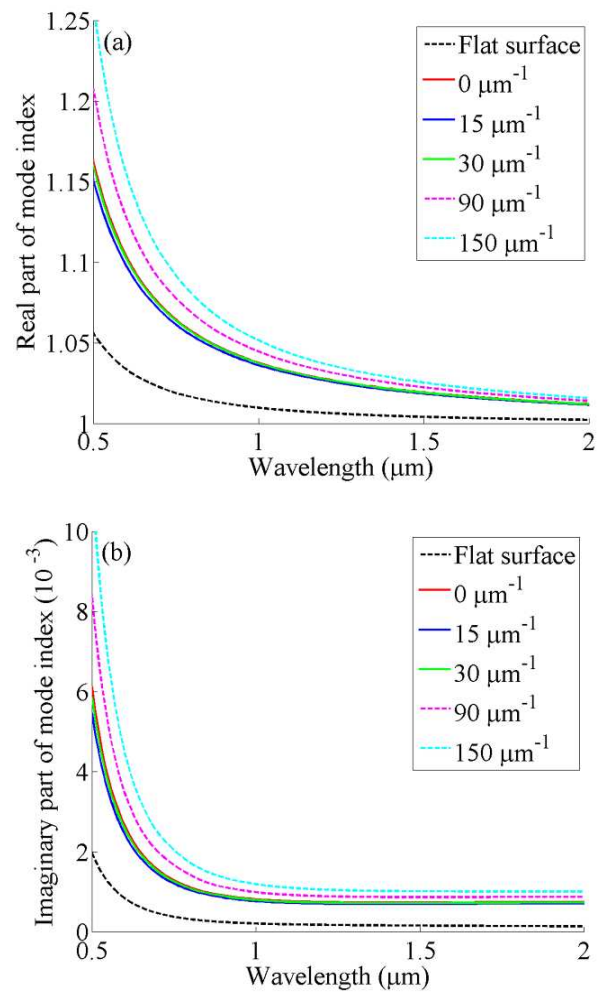

Fig. 5. (a) Real part and (b) imaginary part of the mode index versus the wavelength for different curvatures and the comparison with those of the silver-air flat plasmon mode index.

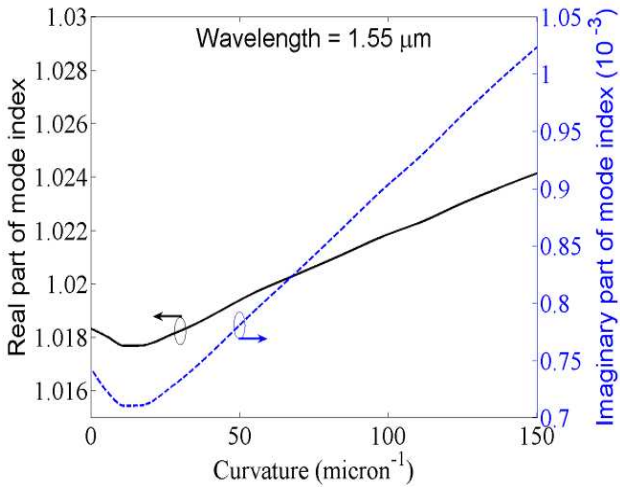

Fig. 6. Real and imaginary part of the mode index versus the curvature at 1.55 $\mu m$ wavelength.

In general, the propagation of surface plasmon modes can be characterized by a complex wave vector of the form $\beta_z = n_{eff} k_0 = \beta - j\alpha$, along the $z$-direction, where $\beta$ is the phase propagation constant of the mode, and $\alpha$ is the attenuation constant. The propagation distance is defined as the distance where the mode intensity attenuates to $1/e$ of its initial value, i.e. $L_p = 1/(2\alpha)$. The propagation distances are calculated for the elliptical nanoridge plasmon waveguide at different curvatures and different free space wavelengths. The results are shown in Fig. 7 (a) and (b). It can be seen that at any given wavelength as the curvature increases, the propagation distance first increases, and then reaches plateau for curvature between 9 and 18 $\mu m^{-1}$, after which it is seen to decrease rather rapidly. As the wavelength increases, the propagation distance also increases. As discussed in the Ref. [26], for the wedge plasmon mode, the propagation loss increases dramatically as the wedge angle decreases. When the curvature is between 9 and 18 $\mu m^{-1}$, it appears that the mode energy is distributed more evenly along the entire ridge, and the associated nanoridge plasmon mode can be considered as an infinite number of coupled large-angle wedge plasmon modes. Within this curvature range, these coupled wedge plasmon modes of the elliptical nanoridge have smaller loss, so the nanoridge plasmon mode has the longest propagation distance.

Outside the metal ridge in the surrounding dielectric medium, the transverse component of the complex wave vector is given by $\beta_\perp = \gamma - j\delta$, where $\gamma$ and $\delta$ describe the field oscillation and decay in the transverse direction, respectively. It follows from Maxwell's equations that the complex propagation constants in the propagation direction and the transverse directions are related as:

$$(\beta - j\alpha)^2 + (\gamma - j\delta)^2 = \varepsilon_d k_o^2 \quad (1)$$

where $k_0$ is the mode propagation constant in the free space, and $\varepsilon_d$ is the dielectric constant of the surrounding dielectric. Solving (1), we can obtain $\gamma$ and $\delta$, which can be used to define the mode size as $1/2\delta + 1/2\delta = 1/\delta$. The mode size of the elliptical nanoridge waveguide versus the free space wavelength and the ellptical curvature is shown in Fig. 8(a), while Fig. 8(b) shows the mode size versus the free space wavelength at various elliptical curvatures. Here is seen that the mode size increases, as the wavelength increases. Once again, for the curvature smaller than 9 $\mu m^{-1}$, the mode size slightly increases, but when the curvature increases higher than 18 $\mu m^{-1}$, the mode size decreases rapidly.

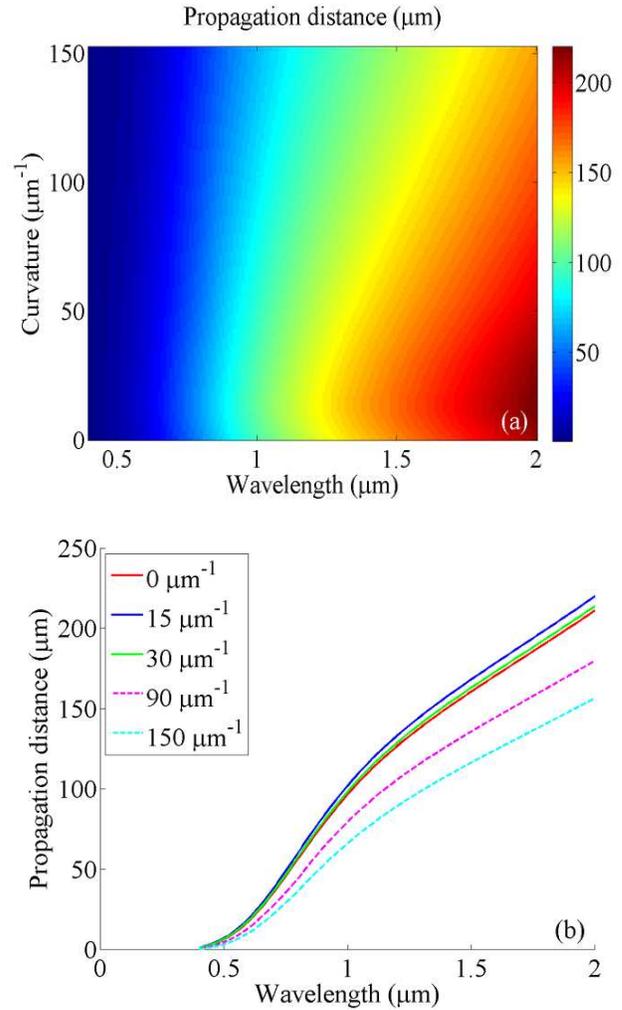

Fig. 7. (a) Propagation distance versus the wavelength and the curvature. (b) Propagation distance of the elliptical nanoridge mode versus the free space wavelength for different curvatures.

While it is always desirable to obtain surface plasmon waveguides with tight confinement and low attenuation, the reality is that there is always a trade-off between the propagation attenuation and the mode confinement [12, 59]. While tight mode confinement is the merit, attenuation is the cost. Figure-of-merits have been proposed to characterize this trade-off between attenuation and mode confinement [60, 61]. Here, we define the figure-of-merit of the nanoridge plasmon waveguide as the ratio of the propagation distance over the mode size:

$$FoM = (1/2\alpha)/(1/\delta) = \delta/2\alpha. \quad (2)$$

We calculated the figure-of-merit versus the wavelength and the curvature. The results are shown in Fig. 9(a). Fig. 9(b) shows the figure-of-merit versus the wavelength for several different curvatures. It can be seen from Fig. 9(a) and (b) that the figure-of-merit reaches a maximum at the 1.05 $\mu m$ wavelength for all curvatures considered in this study. It also can be seen that the magnitude of the figure-of-merit peak is related to the curvature, with values between 9 and 18 $\mu m^{-1}$ producing figure-of-merits that are larger than those obtained for even the zero curvature (the flat-top nanoridge).

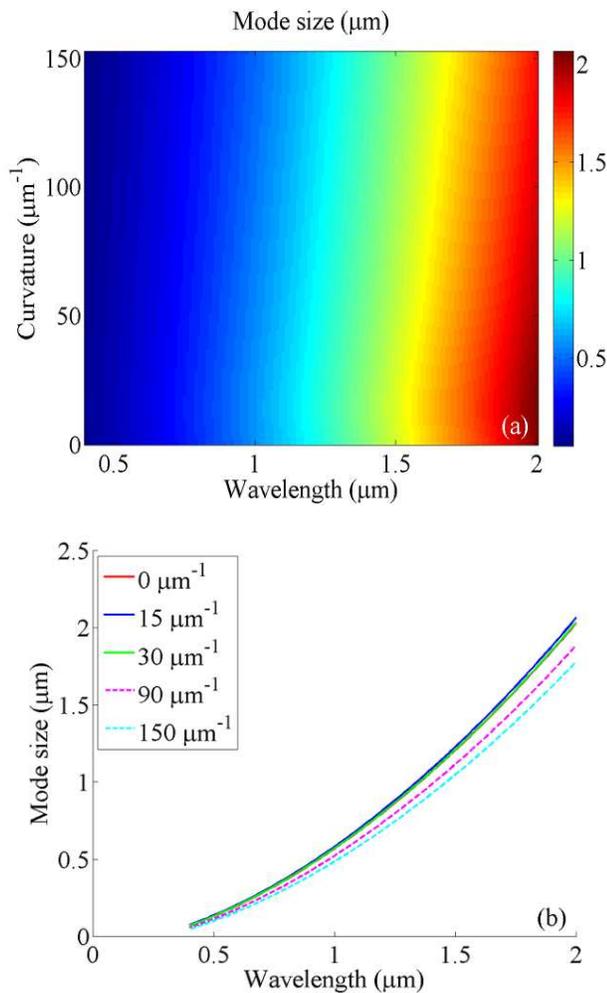

Fig. 8. (a) Mode size versus the wavelength and curvature. (b) Mode size of the elliptical nanoridge plasmon mode versus the free space wavelength for different curvatures.

We specifically calculated the propagation distance and the figure-of-merit for the 120 $nm$ wide elliptical nanoridge waveguide versus the curvature at the wavelength of 1.55 $\mu m$. The results are shown in Fig. 10. As the curvature increases, both the propagation distance and the figure-of-merit increase when the curvature is less than 9 $\mu m^{-1}$, and reach a plateau when the curvature is between 9 and 18 $\mu m^{-1}$, after which both are seen to decrease dramatically. It is concluded that for the 120 $nm$ wide elliptical cross section silver nanoridge waveguides discussed here, the best performance can be obtained with a curvature between 9 and 18 $\mu m^{-1}$ at all wavelengths, with the ultimate performance being obtained when the excitation wavelength is roughly 1.05 $\mu m$.

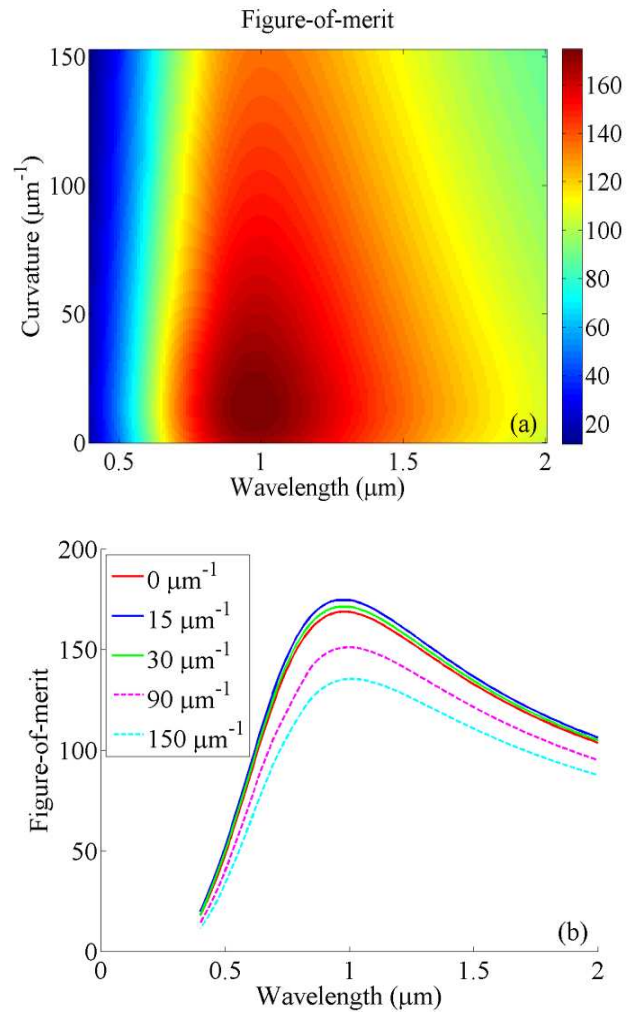

Fig. 9. (a) Figure-of-merit versus the wavelength and the curvature. (b) Figure-of-merit versus the wavelength for several curvatures.

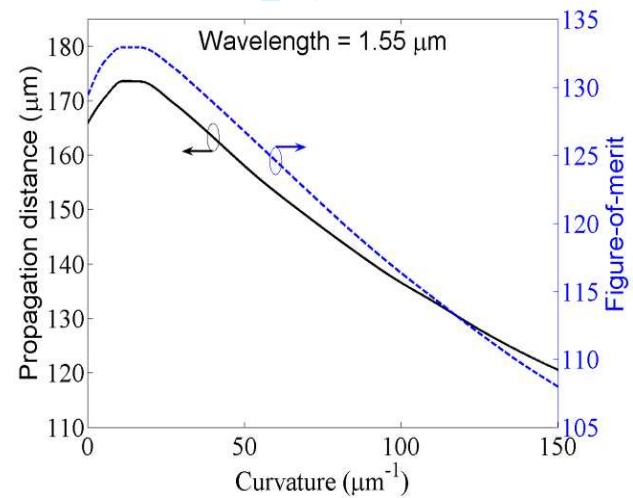

Fig. 10. Propagation distance and figure-of-merit of elliptical nanoridge waveguide mode versus the curvature at 1.55 $\mu m$ wavelength.

We also have calculated the propagation distance and the figure-of-merit for different elliptical height and nanoridge width at the 1.55 $\mu m$ wavelength. The results are shown in Fig. 11. The horizontal axis is the ellipse height of the nanoridge from 0 to 1000 nm. The vertical axis is the width of the nanoridge from 50 to 500 nm. It can be seen that as the height of the elliptical ridge increases, the propagation distance and the figure-of-merit increase first, and then decrease. The white dashed lines in the Fig. 11 indicate the elliptical cross sections which become semi-circles. The propagation distance and figure-of-merit reach the maximal values along this line. This is because the plasmon mode energy is more uniformly distributed over the entire ridge, when the cross section of the nanoridge waveguide becomes a semi-circle.

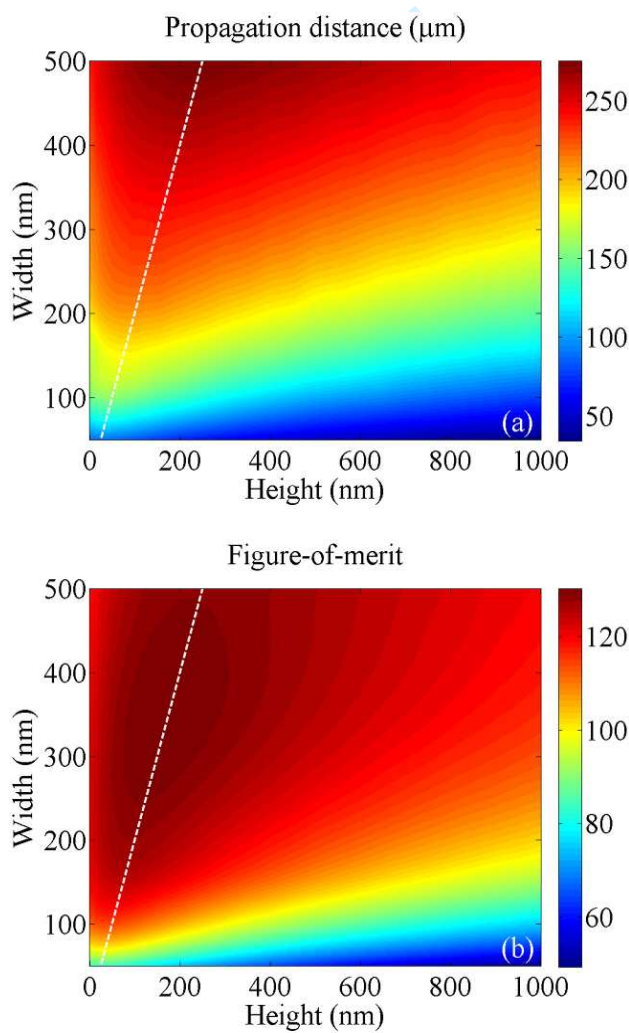

Fig. 11 Propagation distance and figure-of-merit versus the height and the width at 1.55 $\mu m$ wavelength.

III. CONCLUSIONS

We investigated the propagating plasmon modes supported by elliptical cross section silver nanoridges. We calculated the dispersion curves, propagation distances, mode sizes, and the figure-of-merits of these nanoridge plasmon modes. It is found that as the curvature of the elliptical nanoridge increases, both the propagation distance and the figure-of-merit first increase, reach a plateau, and then decrease dramatically. When the curvature falls between 9 and 18 $\mu m^{-1}$ for 120 $nm$ wide silver elliptical ridges, the ridge plasmon modes have the longest propagation distance and the highest figure-of-merit. The elliptical nanoridge waveguides with the curvature between 9 and 18 $\mu m^{-1}$ have better performance in terms of the propagation distance and figure-of-merit than the flat-top nanoridge waveguides investigated previously [53]. It is found that the semicircular cross section nanoridge waveguides, which give more uniformly mode energy distribution over the entire ridge surfaces, represent an optimal trade-off between mode confinement and propagation distance. The elliptical cross section nanoridge waveguides investigated in this paper are very desirable for applications that require large surface interaction area, such as in chemical and biological sensing.

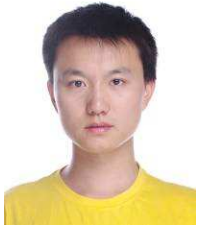

**Zeyu Pan** received his B.S. degree in Optical Information Science & Technology from Nanjing University of Science and Technology, Nanjing, China, in 2010. He is currently pursuing his MS degree at the University of Alabama in Huntsville, Huntsville, AL, USA. His research interests include the surface plasmon waveguide, extraordinary optical transmission, and electromagnetic algorithms.

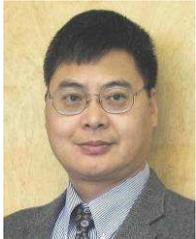

**Junpeng Guo** is currently Associate Professor of Electrical Engineering and Associate Professor of Optical Science and Engineering at the University of Alabama in Huntsville. He earned a Ph.D. degree in electrical engineering and also a MS degree in nuclear plasma engineering from the University of Illinois at Urbana-Champaign. After he graduated from the Illinois, he started to work as a research scientist with the former Rockwell Science Center in Thousand Oaks, California and later as a technical staff member with the Sandia National Laboratories in Albuquerque, New Mexico. Dr. Guo worked in numerous research areas in photonics, optics and lasers. Most notably, he fabricated the first thin film micro-polarizer array for polarization imaging and 3D display. His current research interests are plasmonics, nanophotonics, and metamaterials.

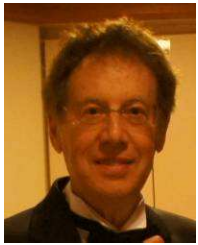

**Richard Soref** received a Ph.D. degree in electrical engineering from Stanford University, Stanford, California in 1963. He is currently a Research Professor at the University of Massachusetts at Boston. His research career in electro-optics spans 48 years. In 2007, he received the Lifetime Achievement Award from the IEEE Photonics Society's International Conference on Group IV Photonics. He is a Life Fellow of IEEE and a Fellow of OSA, AFRL and the Institute of Physics UK.

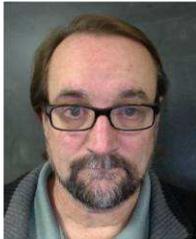

**Walter Buchwald** is currently the Director of Research for Solid State Scientific Corporation, Nashua, New Hampshire. He holds graduate degrees in both Physics and Electrical Engineering earning his Ph.D. from Rutgers University. He has held a variety of technical positions investigating semiconductor materials, devices, and advanced fabrication techniques at the Army Research Laboratory, the Raytheon Advanced Device Center, Axsun Technologies and most recently with the Air Force Research Laboratory where he initiated programs in plasmonics and IR integrated quantum optics. He has over 90 publications and holds two patents and is a graduate faculty scholar with the Physics Department of the University of Central Florida.